# Deconstructing Student Perceptions of Generative AI (GenAI) through an Expectancy Value Theory (EVT)-based Instrument

Author: Cecilia KY CHAN*   Wenxin ZHOU

Affiliation: University of Hong Kong
**Email:** Cecilia.Chan@cetl.hku.hk*Corresponding Author
https://orcid.org/0000-0001-6984-6360
**Email:** zhouwx@connect.hku.hk

**Abstract:** This study examines the relationship between student perceptions and their intention to use generative AI in higher education. Drawing on Expectancy-Value Theory (EVT), a questionnaire was developed to measure students' knowledge of generative AI, perceived value, and perceived cost. A sample of 405 students participated in the study, and confirmatory factor analysis was used to validate the constructs. The results indicate a strong positive correlation between perceived value and intention to use generative AI, and a weak negative correlation between perceived cost and intention to use. As we continue to explore the implications of generative AI in education and other domains, it is crucial to carefully consider the potential long-term consequences and the ethical dilemmas that may arise from widespread adoption.

**Keywords:** Expectancy-Value Theory (EVT); Validated Instrument; Generative AI; ChatGPT

**Introduction**

The recent launch of ChatGPT (Schulman et al., 2022), an advanced language model based on the Generative Pre-trained Transformer (GPT) architecture, has generated significant interest and excitement in both academic and industry circles (Agrawal et al., 2022; Chui et al., 2022; Cotton et al., 2023; Mucharraz y Cano et al., 2023). With its impressive capabilities to generate coherent and contextually appropriate responses that closely mimic human-like communication, ChatGPT has the potential to become a game changer in students' lives, influencing various aspects of their personal, social and professional experiences.

The increasing prevalence of artificial intelligence (AI) in various industries has led to an unprecedented surge in the demand for AI-related skills and knowledge. Generative AI(GenAI), a subset of AI that focuses on generating new content, has shown tremendous potential in applications across numerous domains, revolutionizing the way humans interact with technology and solve complex problems (Russell & Norvig, 2016). In the field of healthcare, AI has been employed in the development of predictive models, diagnosis, and treatment planning, leading to improved patient outcomes (Topol, 2019). In the realm of finance, AI-powered algorithms have facilitated more accurate risk assessment, fraud detection, and algorithmic trading (Königstorfer & Thalmann, 2020). Furthermore, AI has transformed the transportation industry through advancements in autonomous vehicles and traffic management systems (Iyer, 2021). And of course, GenAI has penetrated into the education industry, providing personalized learning experiences, tailoring instructional content to individual students' needs and abilities (Chassignol et al., 2018; Crompton & Burke, 2023). Moreover, it can facilitate collaboration and peer interaction by generating context-aware prompts and responses, fostering a dynamic learning environment that fosters engagement and deeper understanding (Zawacki-Richter et al., 2019). The increasing integration of AI into various industries highlights the growing importance for the future generation to understand and harness its potential and challenges to drive future innovation (Mnih et al., 2015). As educational institutions adapt to this changing landscape, it is crucial to understand students' intentions to use GenAI tools, as it could influence curriculum design, resource allocation, and the future workforce.

The use of GenAI technologies in teaching and learning within higher education is essential to ensure students are adequately prepared for their personal and professional pursuits in this fast-pacing world. However, little is known about students' experience and perceptions of these technologies and their intention to use them. Understanding students' perceptions and intentions is crucial for educators and policymakers to make informed decisions about the integration of GenAI technologies into higher education. Therefore, this study aims to explore students' intention to use GenAI using expectancy-value theory by examining the role of knowledge of and familiarity, perceived value, and cost. The research questions for this study are

1. Is there a correlation between students' knowledge of and familiarity with GenAI and their intention to use GenAI?
2. Is there a correlation between students' perceived value of using GenAI and their intention to use AI?
3. Is there a correlation between students' perceived cost of using GenAI and their intention to use AI?

**Literature on student's perception of AI and GenAI**

Research on student perceptions of AI, particularly GenAI, is scarce. Studies reviewed focus on students' perceptions of AI in various aspects of education, ranging from AI teaching assistants to specific applications such as ChatGPT. Students' perceptions of AI teaching assistants in the United States were investigated using the Technology Acceptance Model (TAM) as a theoretical framework to investigate the perceived usefulness and ease of communication with AI teaching assistants (Kim et al., 2020). TAM posits that the perceived usefulness (PU) and perceived ease of use (PEOU) of a technology are key determinants of its acceptance and use (Abdullah & Ward, 2016; Davis, 1989). The study included 321 college students, and the findings suggest that perceived usefulness and ease of communication with an AI teaching assistant positively predict favorable attitudes, which consequently leads to stronger intention to adopt AI teaching assistant-based education. Students who perceived positively with AI teaching assistants mentioned an increase in efficiency and convenience in online education. However, some students also expressed concerns about the lack of human interaction and the potential for errors or technical glitches.

Zou et al. (2020) employed a sequential explanatory mixed-methods design, which comprised of a survey assessing student perceptions of their current usage and effectiveness of AI-English Language Learning apps for speaking skills enhancement, followed by qualitative interviews to elucidate and interpret the findings from the questionnaire. The sample included 113 Year 1 and Year 2 English for Academic Purposes (EAP) students from an English-speaking university in China. The primary findings reveal that participants expressed positive opinions regarding AI technology's role in developing speaking skills, albeit with certain limitations, such as the absence of personalization and feedback. The potential implications of this study suggest that AI technology may serve as a valuable tool for supporting EAP students in improving their speaking skills.

Haensch et al. (2023) analyzed TikTok the social media content to better understand how students perceive and use ChatGPT. The findings suggest that students are interested in using ChatGPT for various tasks, but there is also a concern about its potential impact on academic integrity. The study highlights the need for educators to consider how they incorporate or regulate AI technologies like ChatGPT in universities to raise awareness among students about ethical considerations when using AI technologies.

A recent study in India (Kumar & Raman, 2022) surveyed 682 students enrolled in full-time business management programmes to gather their opinions on the usage of AI in various aspects of higher education, including the teaching learning process, admission process, placement process, and administrative process. Students generally had positive perceptions of AI usage in higher education, particularly in administrative and admission processes. However, they were more hesitant about AI being used as a partial replacement for faculty members in the teaching-learning process. The study also found that students' prior exposure to AI influenced their perceptions.

Bonsu and Baffour-Koduah (2023) explored the perceptions and intentions of Ghanaian higher education students towards using ChatGPT, using a mixed-method approach guided by the Technology Acceptance Model (TAM) with a sample size of 107 students. The study found that although there was no significant relationship between students' perceptions and their intention to use ChatGPT, students expressed the intention to use and supported its adoption in education, given their positive experiences. Social media was identified as a key source of students' knowledge about ChatGPT, and they perceived more advantages than disadvantages of using it in higher education.

Students' intention to use AI-driven language models like ChatGPT in India was also explored by Raman et al. (2023). This study, framed by Rogers' Perceived Theory of Attributes and based on Expectancy-Value Theory (EVT), aimed to explore the factors that determine university students' intentions to use ChatGPT in higher education. A sample of 288 students participated in the study, which focused on five factors of ChatGPT adoption: Relative Advantage, Compatibility, Ease of Use, Observability, and Trialability. The results revealed that all five factors significantly influenced ChatGPT adoption, with students perceiving it as innovative, compatible, and user-friendly. The potential implications of this study suggest that students are open to using AI-driven language models like ChatGPT in their education and perceive them as valuable resources for independent learning.

In the Netherlands, a study (Abdelwahab et al., 2023) was conducted using a survey completed by 95 students from 27 higher education institutions. The survey questions were categorized into four factors based on a conceptual framework, including students' awareness of AI, teacher's skills in AI teaching, teaching facilities for AI, and the AI curriculum. Respondents were asked to provide their answers using various methods, such as a 5-point Likert scale, ranking, yes or no, or open-response answers. Business students in the Netherlands have expressed concerns regarding their higher education institutions' readiness to prepare them for AI work environments. They feel that the institutions are ill-equipped or have not fully utilized their resources to provide adequate AI-related training. There is an urgent need to update the curriculum and educational facilities for AI work environments and provide more comprehensive training and education on AI-related topics.

A study involved 102 physics students from a German university who evaluated ChatGPT responses to introductory physics questions (Dahlkemper et al., 2023). This study aimed to evaluate how physics students perceive the linguistic quality and scientific accuracy of ChatGPT responses to physics questions. The study used a survey instrument based on the Unified Theory of Acceptance and Use of Technology (UTAUT), and included three statements about students' expectations of AI performance and their attitudes towards AI in general. The items were answered on a 5-point Likert scale. The UTAUT model (Venkatesh et al., 2003) identifies four key factors that influence technology adoption: performance expectancy, effort expectancy, social influence, and facilitating conditions. The key findings suggest that while students generally perceived the linguistic quality of ChatGPT responses positively, they were more critical of the scientific accuracy. Additionally, students who had prior experience with AI were more likely to have positive attitudes towards AI in general.

Several factors have been identified in the literature as influencing students' intention of using GenAI in education. Familiarity with AI technologies, personal innovativeness, and perceived usefulness have been shown to positively affect students' attitudes toward AI (Chassignol et al., 2018). Furthermore, perceived ease of use, which relates to the user-friendliness of AI tools, has been found to be a crucial determinant of students' willingness to adopt AI technologies (Venkatesh et al., 2003).

Based on the above literature, our study intends to employ the expectancy-value theory to investigate the correlation between students' intention to use GenAI and their knowledge, familiarity, perceived value, and cost of GenAI.

**Expectancy-Value Theory and Other Frameworks**

In the previous section, different frameworks such as TAM, UTAUT to explore student perception of AI have been mentioned. For our study, Expectancy-Value Theory (EVT) will be used. EVT posits that individuals' decisions to engage in a particular activity or task are influenced by their expectations of success (expectancy) and the perceived value they attach to that activity (value).

Expectancy refers to an individual's belief in their ability to succeed in a task, while value encompasses several components, such as attainment value, intrinsic value, utility value, and cost (Wigfield & Eccles, 2000). When examining students' intention to use GenAI, this framework can be utilized to address the research questions as follows:

RQ1: Is there a correlation between students' knowledge of and familiarity with GenAI and their intention to use GenAI?

According to the expectancy-value theory, students' knowledge and familiarity with GenAI may influence their expectancy beliefs. The more familiar and knowledgeable students are with the technology such as how they generate outputs, the higher their expectancy beliefs may be, leading to a higher likelihood of adopting GenAI in their learning processes (Wigfield & Eccles, 2000). Previous research has also shown a positive correlation between students' knowledge of technology and their intention to use it such as via the UTAUT model (Venkatesh et al., 2003).

RQ 2: Is there a correlation between students' perceived value of using GenAI and their intention to use AI?

Value is a crucial component of the expectancy-value framework, and it is hypothesized that students who perceive higher value in using GenAI will be more likely to adopt it (Wigfield & Eccles, 2000). Studies have shown that perceived usefulness and perceived ease of use are significant determinants of technology acceptance (Davis, 1989; Teo, 2009; Venkatesh et al., 2003). Maheshwari's (2021) study also highlights the impact of institutional support and perceived enjoyment on students' intentions to continue studying courses online. Specifically, the perceived value components that influence these intentions are:

- Attainment value which refers to the belief that engaging in a behavior will lead to an important goal or outcome. For example, students who believe that using GenAI will improve their academic performance or digital competence may be more likely to use it.
- Intrinsic value refers to the personal enjoyment or satisfaction that a person derives from engaging in a behavior. For example, students who enjoy exploring new technologies or feeling comfortable using GenAI due to the anonymity.
- Utility value refers to the belief that engaging in a behavior will lead to practical benefits, such as improved skills or knowledge. For example, students who believe that using GenAI will help them save time or provide them with unique feedback may be more likely to use it.

*RQ 3: Is there a correlation between students' perceived cost of using GenAI and their intention to use AI?*

Cost refers to the negative aspects or barriers associated with engaging in a particular behavior such as effort, time, undermining the value of education, limiting social interactions, or hindering the development of holistic competencies.

Cost can be seen as a factor that influences an individual's motivation and intention to engage in a behavior. If students perceive the costs of using GenAI to outweigh its benefits, they may be less likely to adopt the technology. Previous research has shown that perceived barriers, such as cost, can negatively affect students' intentions to use technology in education (Flake et al., 2015; Regmi & Jones, 2020; Stüber, 2018). The expectancy-value framework has been widely used in educational research to examine students' motivation, learning, and achievement (Cheng et al., 2020; Sin et al., 2022).

**Why Expectancy-Value Theory?**

The Expectancy-Value Theory (EVT) is widely used and has been adopted across various domains. It is chosen as the theoretical framework for this study over other models such as the Unified Theory of Acceptance and Use of Technology (UTAUT), Technology Acceptance Model (TAM) and Theory of Planned Behavior (TPB) because EVT specifically focuses on the factors that drive individuals' motivation and decision-making processes related to their choices, goals, and performance, which is a major focus of this study. While other models like UTAUT, TAM and TPB offer valuable insights into technology acceptance and adoption, they do not fully capture the motivational factors that are central to EVT.

EVT is considered more suitable for this study because it takes into account the perceived value and cost associated with using GenAI, which are critical factors in determining students' intentions to use such technology. Moreover, EVT also emphasizes the role of students' knowledge and familiarity with GenAI, which is an essential aspect of this study's research questions.

**Methodology**

The methodology for this study employed a cross-sectional survey design using an online questionnaire to gather data on students' familiarity, knowledge, perceived value, perceived costs, and intention regarding the use of GenAI technologies in teaching and learning at universities in Hong Kong.

The participants' opinions were assessed using 23 five-point Likert scale questions (Q1 as frequency scale; Q2-Q23, with response options ranging from 1-Strongly Disagree to 5-Strongly Agree). This allowed the participants to express their level of agreement or uncertainty on each statement. The study used convenience sampling as its sampling technique, wherein the participants were selected based on their accessibility and willingness to participate. To reach the participants, the questionnaire was distributed through a bulk email sent to students. While this approach may not ensure a representative sample of the target population, it allows for the efficient collection of data from a large group of respondents.

The items in the survey were developed based on EVT, it consisted of four main sections: knowledge of GenAI, perceived value of using GenAI, perceived cost of using GenAI, and intention to use GenAI. Table 1 shows the factors, their corresponding questionnaire items, and the analysis methods used and table 2 shows the survey items.

| Factor | Questionnaire Items | No. of Items | Analysis Methods |
|---|---|---|---|
| Use frequency | Q1 | 1 | Descriptive analysis, Pearson's correlation |
| Students' Knowledge of AI | Q2-Q6 | 5 | Descriptive analysis, CFA, Pearson's correlation |
| Student-Perceived Value of AI | Q7-Q17 | 11 | Descriptive analysis, CFA, Pearson's correlation |
|     Attainment value | Q7-Q10 | 4 | |
|     Intrinsic value | Q11-Q13 | 3 | |
|     Utility value | Q14-Q17 | 4 | |
| Student-Perceived Cost of AI | Q18-Q21 | 4 | Descriptive analysis, CFA, Pearson's correlation |
| Students' Intention to Use AI | Q22-Q23 | 2 | Descriptive analysis, Pearson's correlation |

*Note:* CFA=Confirmatory Factor Analysis; SEM= Structural Equation Modelling; IBM SPSS 27 and IBM AMOS 28 were used to conduct the analyses

Table 1: Summary of Factors and questionnaire items

The analyses were conducted in two stages. The first stage focused on descriptive analyses of the responses to reveal participants' perceptions by comparing mean and standard deviation (Table 2). The second stage involved the validation of each factor as specified in the EVT section (Table 1). Confirmatory Factor Analysis (CFA) was used to validate each factor and investigate students' intention of integrating GenAI technologies in higher education. The survey items and factors are already grounded in the EVT framework (Wigfield & Eccles, 2000), which has been well-established in previous research on technology adoption (Venkatesh et al., 2003). Due to this strong theoretical basis, it was decided to use only CFA to validate without using Exploratory Factor Analysis (EFA), which is more focused on hypothesis testing and confirming the hypothesized factor structure, rather than exploring new and unknown factor structures that EFA would provide.

According to Brown (2006), CFA is a hypothesis-driven method that allows for direct testing of the proposed factor structure, which is derived from EVT and previous research on technology adoption in education. This method will enable us to test the relationships between the survey items and their respective factors, as well as examine the correlations among the factors themselves. For example, in the context of our study, we will be able to assess whether the survey items measuring Students' Knowledge of AI (Q2-Q6) indeed load onto a single factor, as theorized. Similarly, we can investigate the relationships between the factors and determine if they are consistent with EVT (e.g., whether the Student-Perceived Value of AI is positively related to Students' Intention to Use AI).

Moreover, the use of CFA in this study can be considered a parsimonious approach, ensuring that our research findings are concise and easier to interpret. This approach allows us to concentrate on the relationships between the survey items and factors, providing a clear picture of how these items are connected to the broader constructs of the EVT framework (Wigfield & Eccles, 2000).

The analyses in the research were made through IBM SPSS 27 and IBM AMOS 28.

## Results
### Demographics and Descriptive Analysis

The survey study was conducted among students from Hong Kong to explore their perceptions of using GenAI technologies like ChatGPT for teaching and learning in higher education. In total, 405 participants provided valid information for data analyses with an average age of 23.87 years, consisting of 51.4% males (n=208) and 48.6% females (n=197).

Table 2 summarized the means and S.D. results of the survey data (Q1-Q23). The use frequency and familiarity with GenAI technologies among participants varied (never=33.6%; rarely=22.0%; sometimes=28.9%; often=9.6%; always=5.9%) based on Q1 ("*I have used generative AI technologies like ChatGPT*"). With a mean as low as 2.32, Q1 demonstrated that many participants had limited user experience with GenAI by the date the research was conducted.

To better understand Q2-Q33, means were interpreted by referring to the Likert Scale interval recommended by Pimentel (2010), where a point mean falls in the range from 1.00 to 1.80 can be regarded as strongly disagree, 1.81 to 2.60 as disagree, 2.61 to 3.40 as neutral, 3.41 to 4.20 as agree, and 4.21 to 5.00 as strongly agree. Thus, at between 3.40 and 2.61, Q11 and Q10 tended to be neutral, while the rest of the items showed an overall agree perception.

|     |     | | Five-point scale | |
| --- | --- | --- | --- | --- |
|     |     | n | Mean | S.D. |
| **Q1** | I have used generative AI technologies like ChatGPT. | 405 | 2.32 | 1.201 |
| **Q2** | I understand generative AI technologies like ChatGPT have limitations in their ability to handle complex tasks. | 399 | 4.15 | 0.823 |
| **Q3** | I understand generative AI technologies like ChatGPT can generate output that is factually inaccurate. | 391 | 4.10 | 0.846 |
| **Q4** | I understand generative AI technologies like ChatGPT can generate output that is out of context or inappropriate. | 395 | 4.04 | 0.832 |
| Q5 | I understand generative AI technologies like ChatGPT can exhibit biases and unfairness in their output. | 385 | 3.93 | 0.917 |
| Q6 | I understand generative AI technologies like ChatGPT have limited emotional intelligence and empathy, which can lead to output that is insensitive or inappropriate. | 387 | 3.89 | 0.969 |
| Q7 | Students must learn how to use generative AI technologies well for their career. | 395 | 4.06 | 0.950 |
| Q8 | I believe generative AI technologies such as ChatGPT can improve my digital competence. | 387 | 3.71 | 0.958 |
| Q9 | I believe generative AI technologies such as ChatGPT can improve my overall academic performance. | 372 | 3.48 | 0.981 |
| Q10 | I think generative AI technologies such as ChatGPT can help me become a better writer. | 387 | 3.32 | 1.161 |
| Q11 | I can ask questions to generative AI technologies such as ChatGPT that I would otherwise not voice out to my teacher. | 390 | 3.38 | 1.095 |
| Q12 | Generative AI technologies such as ChatGPT will not judge me, so I feel comfortable with it. | 386 | 3.53 | 1.059 |
| Q13 | I think AI technologies such as ChatGPT is a great tool for student support services due to anonymity. | 382 | 3.78 | 0.988 |
| Q14 | I believe generative AI technologies such as ChatGPT can help me save time. | 396 | 4.20 | 0.820 |
| Q15 | I believe AI technologies such as ChatGPT can provide me with unique insights and perspectives that I may not have thought of myself. | 390 | 3.74 | 1.074 |
| Q16 | I think AI technologies such as ChatGPT can provide me with personalized and immediate feedback and suggestions for my assignments. | 386 | 3.61 | 1.056 |
| Q17 | I think AI technologies such as ChatGPT is a great tool as it is available 24/7. | 394 | 4.13 | 0.826 |
| Q18 | Using generative AI technologies such as ChatGPT to complete assignments undermines the value of a university education. | 393 | 3.15 | 1.172 |
| Q19 | Generative AI technologies such as ChatGPT will limit my opportunities to interact with others and socialize while completing coursework. | 389 | 3.05 | 1.197 |
| Q20 | Generative AI technologies such as ChatGPT will hinder my development of generic or transferable skills such as teamwork, problem-solving, and leadership skills. | 388 | 3.11 | 1.225 |
| Q21 | I can become over-reliant on generative AI technologies. | 382 | 2.86 | 1.125 |
| Q22 | The integration of generative AI technologies like ChatGPT in higher education will have a positive impact on teaching and learning in the long run. | 387 | 3.93 | 0.838 |
| Q23 | I envision integrating generative AI technologies like ChatGPT into my teaching and learning practices in the future. | 383 | 3.86 | 1.014 |

Table 2: Descriptive analysis results (N=405)

**Validity and reliability of the scales**

Driven by the theory of EVT, the questionnaire aimed to understand students' perceptions of using GenAI. Three separate CFA tests were conducted to validate the constructs measuring students' knowledge of GenAI, the student-perceived value of using GenAI, and the student-perceived cost of using GenAI.

The results in Table 3 indicated a good model fit regarding students' knowledge, student-perceived value, and student-perceived cost with $\chi^2/df$ ratio all less than 3, the root mean square error of approximation (RMSEA) all lower than 0.07 (Steiger, 2007) and standardised root mean square residual (SRMR) lower than 0.080 (Hu & Bentler,1999). The comparative fit index (CFI) and Tucker–Lewis index (TLI) of knowledge and perceived cost were all higher than 0.95 (Hu & Bentler,1999). Though TLI of perceived value was less than 0.95 yet still higher than 0.90 and its CFI was higher than 0.95.

| | $\chi^2$ | df | $\chi^2/df$ | RMSEA | 90%CI | CFI | TLI | SRMR |
|---|---|---|---|---|---|---|---|---|
| *Knowledge* | 12.827 | 5 | 2.565 | 0.062 | [0.020,0.105] | 0.989 | 0.979 | 0.0261 |

| | | | | | | | |
|---|---|---|---|---|---|---|---|
| *Perceived value* | 111.107 | 41 | 2.710 | 0.065 | [0.051,0.080] | 0.955 | 0.940 | 0.0395 |
| *Perceived cost* | 4.902 | 2 | 2.451 | 0.060 | [0.000,0.130] | 0.994 | 0.983 | 0.0194 |

*Notes*: RMSEA = root mean square error of approximation; CI = confidence interval; CFI = comparative fit index; TLI = Tucker–Lewis index; SRMR = standardised root mean square residual.

Table 3: CFA results (N=405)

To further measure the reliability of the constructs, Cronbach's alpha coefficient was calculated to test the internal consistency. Cronbach's alphas for the 5 knowledge, 11 perceived value, and 4 perceived cost items were 0.824, 0.874, and 0.759 respectively. As summarized in Table 4, the Cronbach alpha values were all greater than 0.7, indicating an acceptable internal consistency within the three scales.

| | N of items | Mean | Variance | S.D. | Cronbach's Alpha |
|---|---|---|---|---|---|
| Knowledge | 5 | 20.10 | 11.092 | 3.330 | 0.824 |
| Perceived value | 11 | 40.83 | 52.105 | 7.218 | 0.874 |
| Perceived cost | 4 | 12.19 | 12.544 | 3.542 | 0.759 |

Table 4: Cronbach Alpha Coefficient results (N=405)

**Correlations among the variables**

The correlations between students' knowledge, perceived value, and the perceived cost of using GenAI were analyzed using bivariate correlation with Pearson's correlation coefficient (*r*). Pearson's correlation coefficient is used to measure the linear relationship between factors derived from EVT and students' intention to use GenAI in higher education for a sample of 405 participants. The results (see Table 5) suggested a relatively high and positive correlation between student-perceived value (*r*= 0.603, p<0.001) and students' intention to use. The three subscales—attainment value (r= 0.587, p<0.001), intrinsic value (r= 0.456, p<0.001), and utility value (r= 0.5502, p<0.001)—were also positively correlated with students' intention to use GenAI.

The correlations between students' knowledge (*r*= 0.179), student-perceived cost (*r*= -0.301) of GenAI, and their intention to use were more moderate but still significant. Compared with knowledge, the connection between the student-perceived cost of using generating AI and students' intention to use GenAI was stronger, though in a negative way.

| | Pearson Correlation | Sig. (2-tailed) | 95% Confidence Intervals (2-tailed)[a] | |
|---|---|---|---|---|
| | | | Lower | Upper |
| Knowledge – Intention to use | 0.179 | <0.001 | 0.083 | 0.271 |
| Perceived value – Intention to use | 0.603 | <0.001 | 0.537 | 0.661 |
| • Attainment value – Intention to use | 0.587 | <0.001 | 0.519 | 0.647 |
| • Intrinsic value – Intention to use | 0.456 | <0.001 | 0.375 | 0.530 |
| • Utility value – Intention to use | 0.502 | <0.001 | 0.425 | 0.571 |
| Perceived Cost – Intention to use | -0.301 | <0.001 | -0.387 | -0.209 |

a. Estimation is based on Fisher's r-to-z transformation.

Table 5: Correlation analysis results (Pearson's correlation coefficient) (N=405)

**Discussion**

The findings from table 5 demonstrates that EVT-related factors, such as knowledge, perceived value (including attainment, intrinsic, and utility values), and perceived cost, are all significantly correlated with students' intention to use GenAI in higher education. The perceived value has the strongest positive correlation with intention to use, while the perceived cost has a weak negative correlation.

The student-perceived value of GenAI emerged as the most significant factor influencing their intention to utilize such technologies in an educational context. The majority of participants acknowledged the potential advantages of GenAI in the workplace and its capacity to enhance learning outcomes, encompassing the improvement of academic performance and the development of digital competence. Moreover, students identified utility value in aspects such as increased efficiency, provision of personalized and immediate feedback, and facilitation of idea generation.

The correlation analysis between students' knowledge of GenAI and their intention to use it revealed a statistically significant, albeit weak relationship. The findings suggest that while it is important to provide students with basic knowledge about GenAI, such as its definition, limitations, and benefits, this alone is not sufficient to foster their intention to use it. Instead, it may be more critical to foster AI literacy, offer guidance on how to effectively utilize the technology and transform its advantages into valuable outcomes in learning and working contexts. Therefore, institutions should aim to enhance students' understanding of the practical applications of GenAI and provide them with training on how to utilize the technology effectively to achieve desired outcomes.

The findings also suggest that reducing the perceived costs associated with the use of GenAI could potentially increase students' intention to use it. The costs extend beyond monetary considerations to encompass issues such as academic integrity, privacy, and security. Despite GenAI's potential advantages, students voiced concerns about its influence on higher education's value and the cultivation of holistic competencies. To tackle these apprehensions, the study advises fostering social and experiential learning (Chan, 2022) as well as promoting interpersonal interactions within higher education environments.

**Implications**

The implications of this study, which employed a validated instrument grounded in Expectancy-Value Theory (EVT) to evaluate student perceptions of GenAI in higher education, are multifaceted and have far-reaching consequences for researchers, educators, and educational institutions alike.

First and foremost, this study highlights the significance of EVT-related factors, such as knowledge, perceived value, and perceived cost, in shaping students' intention to use GenAI. By identifying these factors and their relationships with intention to use AI, the study provides valuable insights for educators and institutions looking to foster AI adoption in higher education. By emphasizing the potential value of GenAI, addressing concerns related to perceived costs, and enhancing students' knowledge about these technologies, institutions can develop strategies and interventions aimed at promoting positive attitudes towards AI and ultimately improving the learning experience for students.

Second, the study has implications for the design of educational curricula and professional development programs. The findings suggest that institutions should focus not only on providing students with basic knowledge about GenAI but also on fostering AI literacy, offering guidance on effective utilization of AI technologies, and highlighting their practical applications in various learning and working contexts. By incorporating these elements into their curricula, institutions can ensure that students are equipped with the skills and knowledge necessary to make the most of AI in their academic pursuits and future careers.

Moreover, the study's results can inform the development of targeted interventions for different student groups. As individual differences may play a role in shaping perceptions of GenAI, understanding the specific concerns and motivations of various student populations can help educators tailor their interventions to better address the needs of their students. For example, by identifying the factors that may be more important for students in specific disciplines or cultural contexts, educators can design interventions that are more likely to be effective in fostering AI adoption among these groups.

Additionally, the development of a validated instrument based on EVT represents a significant contribution to the field. To date, there has been a lack of robust, theoretically grounded instruments to assess students' attitudes towards GenAI adoption, making it challenging to systematically understand the factors that influence their intention to use the technologies. The EVT-based instrument addresses this gap in the literature and provides a strong foundation for future research and practice in this area.

The validated instrument's strong theoretical foundation in EVT ensures that the constructs being measured are relevant and meaningful, which enhances the validity and generalizability of the findings. By adopting EVT, the instrument takes into account both the cognitive (e.g., knowledge) and affective (e.g., value and cost) dimensions of students' perceptions, providing a comprehensive picture of the factors shaping their intention to use GenAI.

In addition to its applicability in evaluating student perceptions, this instrument can be adapted to examine the attitudes of educators towards GenAI in different contexts. For instance, the instrument could be tailored to assess teachers' knowledge of GenAI, their perceptions of its value in the classroom, and the potential costs associated with its implementation. Such information could be invaluable for educational institutions aiming to promote AI adoption among their teaching staff, as it would enable them to address potential barriers and develop targeted professional development opportunities.

Moreover, the instrument's adaptability across different educational contexts allows researchers and practitioners to compare the factors influencing AI adoption among various populations, such as students and educators in different levels, countries or academic disciplines. This can provide valuable insights into the contextual factors that might shape perceptions of GenAI and inform the development of context-specific interventions and policies to support its adoption in higher education.

In conclusion, the development of a validated EVT-based instrument to evaluate student perceptions of GenAI holds significant potential for advancing our understanding of the factors influencing AI adoption in higher education. By providing a theoretically grounded, reliable, and adaptable tool for assessing perceptions, this instrument can facilitate research and practice aimed at fostering positive attitudes towards AI adoption among both students and educators across diverse educational contexts.

**Limitation**

Despite the valuable insights provided by this study, it is important to acknowledge its limitations. Firstly, the sample size was restricted to 405 participants, which may not be fully representative of the larger student population. Additionally, the study was conducted at a single point in time, and thus, it may not account for potential changes in students' attitudes and perceptions towards GenAI over time. The focus on higher education students also limits the generalizability of the findings to other age groups or educational contexts. Furthermore, the study primarily relied on self-reported data, which may be subject to response biases, such as social desirability or recall bias. Lastly, the study did not explore the influence of individual differences, such as cultural background, personal experiences, and learning styles, which could also impact students' perceptions of GenAI.

It is essential to acknowledge that skipping EFA in the validation can have some limitations, such as the potential to overlook alternative factor structures or issues with item loadings (Fabrigar et al., 1999). However, given the strong theoretical basis, prior research, and focus on hypothesis testing, using only CFA in this study can be considered a justifiable decision. To address potential concerns, it may be helpful to consider conducting additional validation studies in the future to further explore the factor structure and psychometric properties of the survey instrument.

Future research should aim to address these limitations by employing larger and more diverse samples, conducting longitudinal studies, and examining the impact of individual differences on the relationship between perceived value, perceived cost, and the intention to use GenAI.

**Conclusion**

In drawing conclusions from this study, an intriguing question emerges: What would be the ultimate threshold (ie.) the equilibrium point between perceived value and cost that determines whether students should use or continue to use GenAI or not?

Furthermore, what if one day, the perceived cost becomes too high, with concerns such as ethics, integrity, and human values being at risk? Would we, as a society, decide to stop using GenAI, or would it be too late to reverse course?

While the correlation analysis sheds light on the relationship between perceived value, perceived cost, and the intention to use GenAI, it does not pinpoint a specific threshold at which students would cease using AI. The strong positive correlation between perceived value and intention to use AI, along with the weak negative correlation between perceived cost and intention to use AI, suggests that as long as the perceived value outweighs the perceived cost, students are more likely to continue using AI technologies.

However, determining an exact point at which students would stop using AI due to the balance of perceived value and cost would require a more detailed analysis, possibly involving a regression model to predict intention to use AI based on the perceived value and cost factors. Additionally, individual differences among students and their unique experiences and priorities may influence their decision-making, making it difficult to pinpoint a universal threshold.

As we continue to explore the applications and implications of GenAI in education and other domains, it is crucial to carefully consider the potential long-term consequences and the ethical dilemmas that may arise from widespread adoption. By weighing these factors, we can make informed decisions about the future of GenAI and ensure that its use aligns with our core values as a society.

**Declarations:**

**Acknowledgements**
The author wishes to thank the students and teachers who participated the survey.

**Conflict of interest**
There is no potential conflict of interest.

**Data availability statement**
Availability of data and material: The datasets used and/or analysed during the current study are available from the corresponding author on reasonable request.